\documentclass[journal]{IEEEtran}

\ifCLASSINFOpdf
\else
   \usepackage[dvips]{graphicx}
\fi
\usepackage{url}

\usepackage{graphicx,balance}
\usepackage{bm}
\usepackage{amssymb}
\usepackage{amsmath}
\usepackage{textcomp}
\usepackage{color}
\usepackage{epstopdf}
\usepackage{subfigure}
\usepackage{booktabs}
\usepackage{enumitem}
\usepackage{algorithm}
\usepackage{algorithmic}
\usepackage{bbm}
\usepackage{makecell}
\usepackage{multirow}

\begin{document}

\title{Corrections to ``Computer Vision Aided mmWave Beam Alignment in V2X Communications"}

\author{Weihua Xu, Feifei Gao, Xiaoming Tao, Jianhua Zhang, and Ahmed Alkhateeb
\thanks{W. Xu and F. Gao are with Institute for Artificial Intelligence, Tsinghua
University (THUAI), Beijing National Research Center for Information
Science and Technology (BNRist), Department of Automation, Tsinghua
University, Beijing, P.R. China, 100084 (email: xwh19@mails.tsinghua.edu.cn,
feifeigao@ieee.org).}
\thanks{X. Tao is with the Department of Electronic Engineering, Tsinghua University, Beijing, P.R. China, 100084 (email: taoxm@tsinghua.edu.cn).}
\thanks{J. Zhang is with the State Key Laboratory of Networking and Switching Technology, Beijing University of Posts
and Telecommunications, Beijing 100876, China (e-mail: jhzhang@bupt.edu.cn).}
\thanks{A. Alkhateeb is with the Department of Electrical and Computer Engineering, University of Texas at Austin, Austin, TX 78712-1687 USA (e-mail:
alkhateeb@asu.edu).}
}

\maketitle

\begin{abstract}
In this document, we revise the results of \cite{WXu} based on more reasonable assumptions regarding data shuffling and parameter setup of deep neural networks (DNNs). Thus, the simulation results can now more reasonably demonstrate the performance of both the proposed and compared beam alignment methods. We revise the simulation steps and make moderate modifications to the design of the vehicle distribution feature (VDF) for the proposed vision based beam alignment when the MS location is available (VBALA). Specifically, we replace the 2D grids of the VDF with 3D grids and utilize the vehicle locations to expand the dimensions of the VDF. Then, we revise the simulation results of Fig. 11, Fig. 12, Fig. 13, Fig. 14, and Fig. 15 in \cite{WXu} to reaffirm the validity of the conclusions.
\end{abstract}

\IEEEpeerreviewmaketitle

\section{Corrections to VBALA Method}
Fig.~1 shows the corrected diagram of VBALA \cite{WXu}. Compared with the 2D grids based VDF in \cite{WXu}, we divide the cubic RSU coverage area into $G_\mathrm{X}\times G_{\mathrm{Y}}\times G_{\mathrm{Z}}$ 3D grids with length $L_{\mathrm{G}}$, width $W_\mathrm{G}$, and height $H_\mathrm{G}$. Moreover, for each 3D grid, we set a local coordinate system (LCS) with $\mathrm{X}_{\mathrm{L}}-\mathrm{Y}_{\mathrm{L}}-\mathrm{Z}_{\mathrm{L}}$ axis, where $\mathrm{X}_{\mathrm{L}}$ axis is parallel to the $\mathrm{X}_{\mathrm{R}}$ axis and the origin is set as a grid vertex that can ensure the grid is in the LCS's first quadrant. Since all vehicles run on the $\mathrm{X}_{\mathrm{R}}-\mathrm{Y}_{\mathrm{R}}$ plane, we set the $\mathrm{Z}_{\mathrm{R}}$-axis center coordinate $z_{\mathrm{R}}^{i,j}$ of the $(i,j)$th vehicle to half of the vehicle height $h_{i,j}$. Then, denote the grid index $(g_\mathrm{X},g_\mathrm{Y},g_\mathrm{Z})$ as $\bm{g}$. Similar to \cite{WXu}, for the vehicles whose center locations are contained in the $\bm{g}$th grid, we obtain their average length, width, height, azimuth, and center coordinates under the corresponding LCS as $l_{\mathrm{ave}}^{\bm{g}}$, $w_{\mathrm{ave}}^{\bm{g}}$, $h_{\mathrm{ave}}^{\bm{g}}$, $\theta_{\mathrm{R}}^{\bm{g}}$, and $(x_{\mathrm{L}}^{\bm{g}},y_{\mathrm{L}}^{\bm{g}},z_{\mathrm{L}}^{\bm{g}})$, respectively.

The corrected VDF is defined as a $G_\mathrm{X}\times G_{\mathrm{Y}}\times G_{\mathrm{Z}}\times 7$ dimensional matrix $\bm{F}_{\mathrm{cor}}$, and the $\bm{g}$th row of $\bm{F}_{\mathrm{cor}}$ is set as $\left[\frac{x_{\mathrm{L}}^{\bm{g}}}{W_\mathrm{G}},\frac{y_{\mathrm{L}}^{\bm{g}}}{L_\mathrm{G}},\frac{z_{\mathrm{L}}^{\bm{g}}}{H_\mathrm{G}},\frac{w_{\mathrm{ave}}^{\bm{g}}}{W_{\mathrm{max}}},\frac{l_{\mathrm{ave}}^{\bm{g}}}{L_{\mathrm{max}}},\frac{h_{\mathrm{ave}}^{\bm{g}}}{H_{\mathrm{max}}},\frac{\theta_{\mathrm{R}}^{\bm{g}}}{2\pi}\right]$. To adapt to $\bm{F}_{\mathrm{cor}}$, the 1D convolution layers of the VDF based beam alignment deep neural
network (VDBAN) in \cite{WXu} are straightforwardly replaced with the 3D convolution layers. All other method steps remain unchanged.

\begin{figure}[t]
\centering
\includegraphics[width=0.5\textwidth]{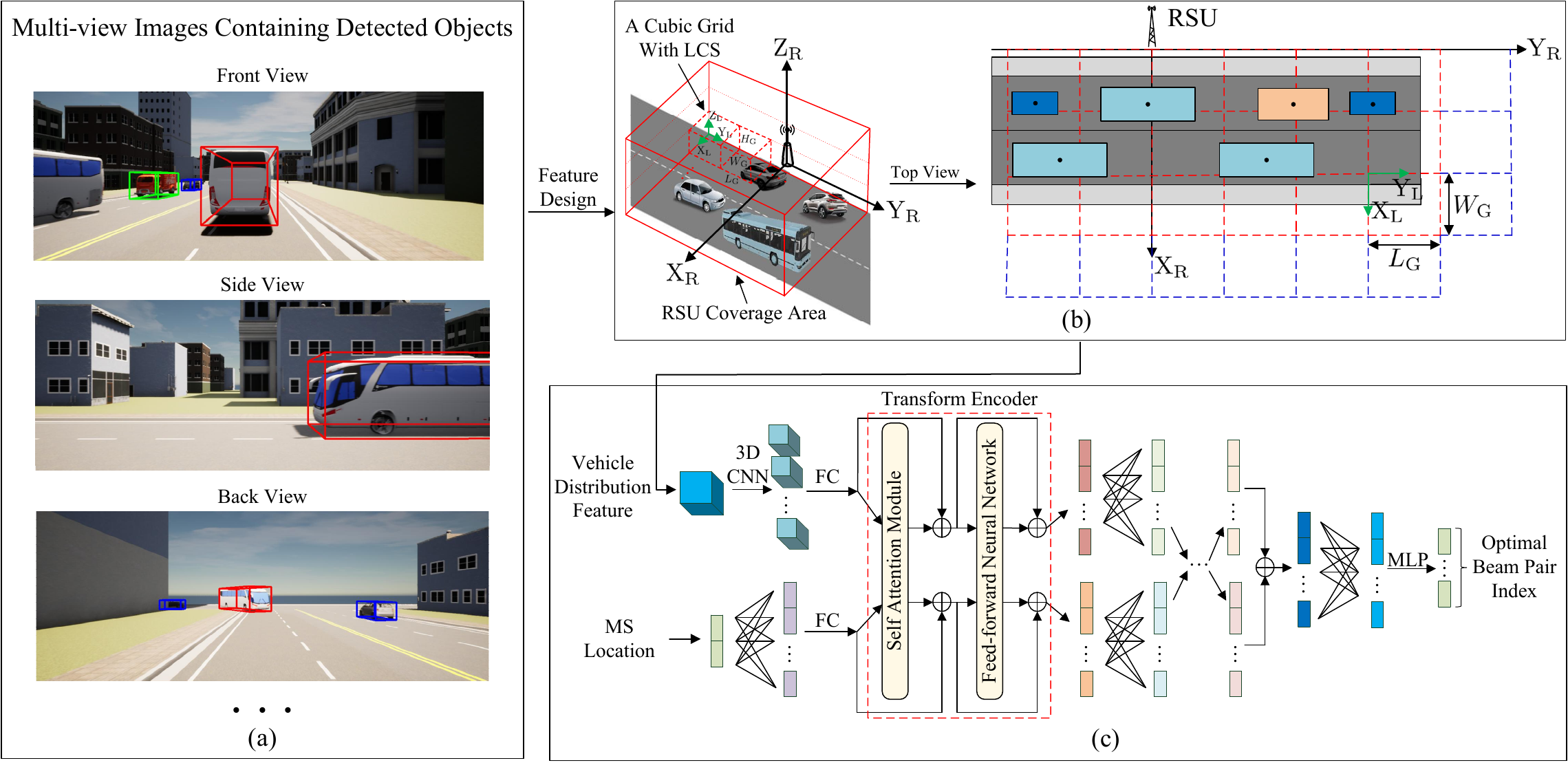}
\caption{The diagram of the corrected VBALA.}
\end{figure}

\section{Corrections to the Simulation Results}
\begin{figure}[t]
\centering
\includegraphics[width=0.5\textwidth]{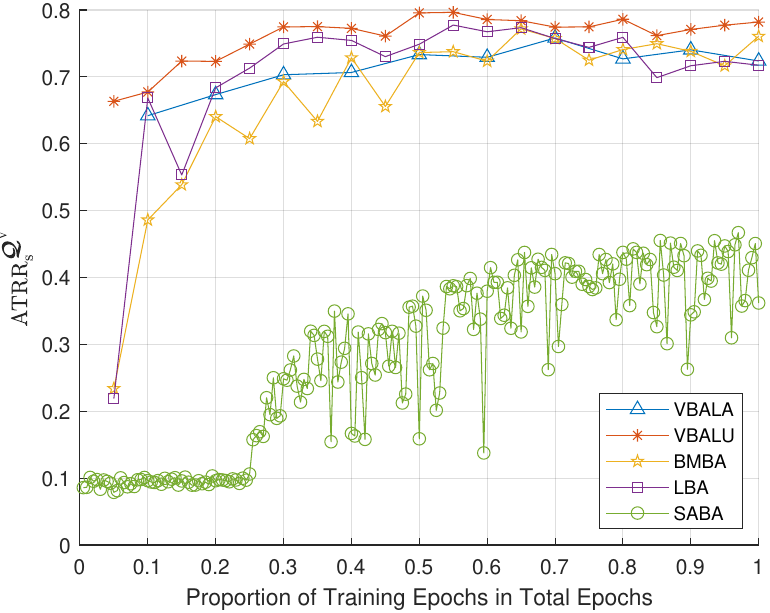}
\caption{${\mathrm{ATRR}}_{\mathrm{s}}^{\bm{\mathcal{Q}}^{\mathrm{V}}}$ achieved by Top-1 beam pair selection with the increase of the number of training epochs.}
\end{figure}

\begin{figure}[t]
\centering
\includegraphics[width=0.5\textwidth]{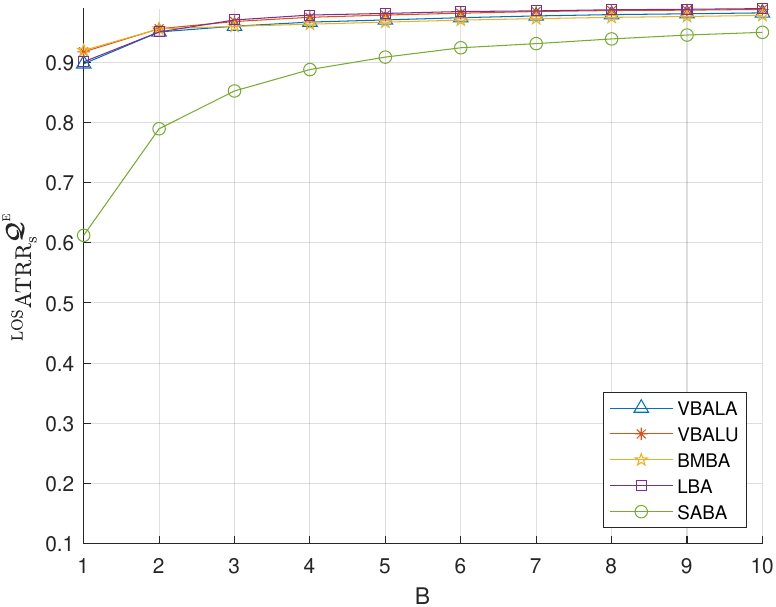}
\caption{$^{\mathrm{LOS}}\mathrm{ATRR}_{\mathrm{s}}^{\bm{\mathcal{Q}}^{\mathrm{E}}}$ for Top-B beam pair selection. The number of LOS test samples are 1930, which is 58\% of total test samples.}
\end{figure}

\renewcommand{\thefigure}{5}
\begin{figure}
\centering
\includegraphics[width=0.5\textwidth]{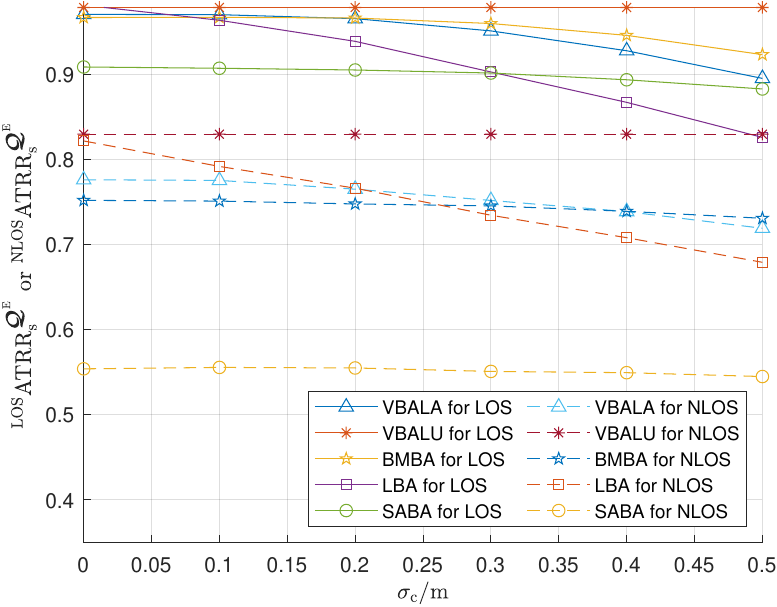}
\caption{$^{\mathrm{LOS}}\mathrm{ATRR}_{\mathrm{s}}^{\bm{\mathcal{Q}}^{\mathrm{E}}}$ and $^{\mathrm{NLOS}}\mathrm{ATRR}_{\mathrm{s}}^{\bm{\mathcal{Q}}^{\mathrm{E}}}$ for Top-5 beam pair selection with different location error $\mathcal{N}(0, \sigma_{\mathrm{c}}^2)$.}
\end{figure}

\renewcommand{\thefigure}{4}
\begin{figure}[t]
\centering
\includegraphics[width=0.5\textwidth]{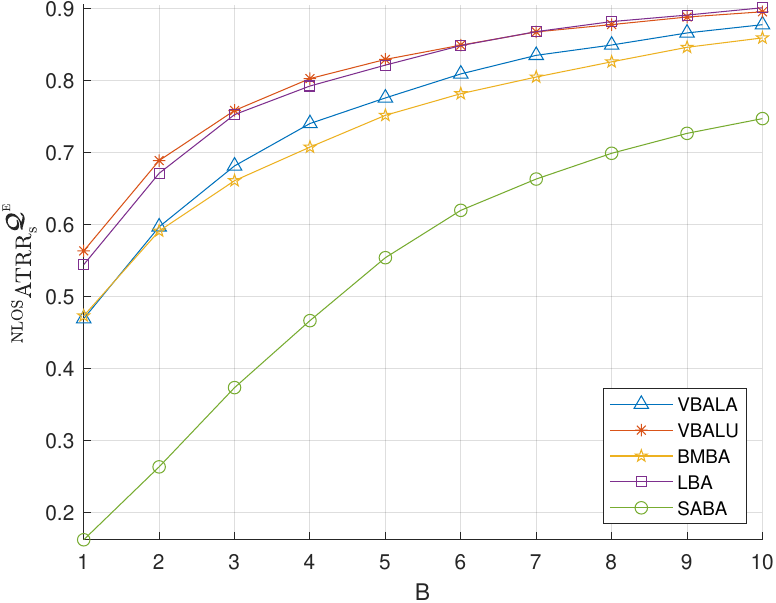}
\caption{$^{\mathrm{NLOS}}\mathrm{ATRR}_{\mathrm{s}}^{\bm{\mathcal{Q}}^{\mathrm{E}}}$ for Top-B beam pair selection. The number of NLOS test samples are 1410, which is 42\% of total test samples.}
\end{figure}

\renewcommand{\thefigure}{6}
\begin{figure}
\centering
\includegraphics[width=0.5\textwidth]{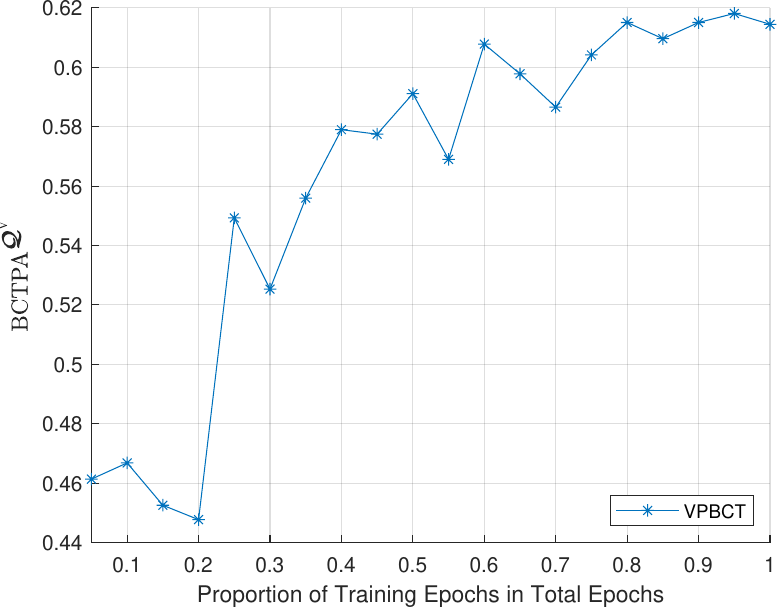}
\caption{$\mathrm{BCTPA}^{\bm{\mathcal{Q}}_{\mathrm{V}}}$ with the increase of the number of training epochs.}
\end{figure}

In the simulation of \cite{WXu}, the dataset for each DNN of beam alignment is only shuffled once at the initialization stage of the training phase. Thus, the performance of the proposed and compared methods is not be reasonably demonstrated. To address the simulation issue, we modify the training approach to the standard mode with dataset shuffle at every epoch. Moreover, we adjust some parameters of the DNNs' structure to compare the performance of all methods more reasonably. Specifically, for the corrected VBALA, $W_{\mathrm{G}}$, $L_{\mathrm{G}}$, and $H_{\mathrm{G}}$ are set as $2\mathrm{m}$, $6\mathrm{m}$, and $1\mathrm{m}$ respectively, and the dimension $G_{\mathrm{X}}\times G_{\mathrm{Y}}\times G_{\mathrm{Z}}$ are set as $10\times6\times6$ respectively. The corrected VDBAN has 10 3D convolution layers with the filter numbers 6, 6, 8, 8, 16, 16, 32, 32, 64, and 64, respectively. The kernel sizes of these 3D convolution layers are (5,5,5), (5,5,5), (5,5,3), (5,5,3), (5,3,3), (5,3,3), (3,3,3), (3,3,3), (3,3,3), and (3,3,3), respectively. The node numbers of the following fully connected (FC) layers are set as 512, 256, 128, and 64, respectively. The final MLP network of the corrected VDBAN has three FC layers with node numbers 512, 1024 and 365, respectively. The corrected VDBAN is trained for 10 epochs. For the DNN of BS's vision based multi-modal beam alignment (BMBA), the feature extraction
subnetwork for images is set as ResNet101V2, which is the same as the DNN of vision based beam alignment when the MS location is unavailable (VBALU). The node numbers of the following FC layers are set as 128, 128, 256, 256, 512, 1024 and 365 to be consistent with that of the corrected VDBAN. The DNN of situational awareness based beam alignment (SABA) \cite{SABA} has 13 FC layers with 128, 128, 256, 256, 512, 1024,  1024, 1024, 1024, 1024, 1024, 1024, and 365 nodes. The DNN of SABA is trained for 200 epochs. For LIDAR based beam alignment (LBA), the LIDAR is set at $1\mathrm{m}$ above the roof center of vehicle. The grid size of the point cloud feature (PCF) is set as $0.1\mathrm{m}\times0.15\mathrm{m}\times0.5\mathrm{m}$, and the grid dimension of PCF is set as $256\times256\times12$. All other simulation settings are not changed. Note that the floating point operations (FLOPs) of the corrected VDBAN is $2.01\times10^{8}$, which is still significantly lower than $1.22\times 10^{10}$ and $1.47\times 10^{10}$ FLOPs, and $1.98\times 10^{10}$ of the DNNs of VBALU, BMBA, and LBA, respectively.

We obtain the corrected simulation results in Fig.~2, Fig.~3, Fig.~4, and Fig.~5, which corresponds to the Fig.~11, Fig.~12, Fig.~13, and Fig.~14, respectively in \cite{WXu}. Compared with the original simulation results in \cite{WXu}, there are only two differences in the conclusions drawn from the corrected simulation results. The first difference lies in the comparison between Fig.~4 and Fig.~13 in \cite{WXu}. It can be seen that LBA outperforms VBALA and achieves similar performance to VBALU. Nevertheless, as shown in Fig.~5, the performance of LBA is susceptible to the location error $\sigma_\mathrm{c}$, while VBALU is unaffected by $\sigma_\mathrm{c}$. Thus, VBALU can still have a significant performance advantage over LBA especially under larger location errors. The second difference lies in the comparison between Fig.~5 and Fig.~14 in \cite{WXu}. It is seen that the decreasing trend of $^{\mathrm{NLOS}}\mathrm{ATRR}_{\mathrm{s}}^{\bm{\mathcal{Q}}^{\mathrm{E}}}$ of BMBA becomes lower than VBALA, which causes $^{\mathrm{NLOS}}\mathrm{ATRR}_{\mathrm{s}}^{\bm{\mathcal{Q}}^{\mathrm{E}}}$ of BMBA can outperform that of VBALA for $\sigma_\mathrm{c}>0.4\mathrm{m}$. Nevertheless, for $\sigma_{\mathrm{c}}\in[0,0.4]\mathrm{m}$, $^{\mathrm{NLOS}}\mathrm{ATRR}_{\mathrm{s}}^{\bm{\mathcal{Q}}^{\mathrm{E}}}$ of VBALA is approximately 1.44\% higher on average than BMBA, while $^{\mathrm{LOS}}\mathrm{ATRR}_{\mathrm{s}}^{\bm{\mathcal{Q}}^{\mathrm{E}}}$ of VBALA is approximately 0.41\% lower on average than BMBA. Thus, for the location error $\sigma_{\mathrm{c}}$ within a certain range, VBALA can still outperform BMBA and has the advantages of low computational complexity and strong environmental generalization ability. For larger $\sigma_{\mathrm{c}}$, VBALU can be adopted to achieve significantly better performance than BMBA. For all other conclusions, their validity can still be demonstrated by the corrected simulation results. The datasets and scripts for the proposed methods are publicly available \cite{github}.

Additionally, due to negligence, Fig.~15 of \cite{WXu} incorrectly provides the validation accuracies $\mathrm{BCTPA}^{\bm{\mathcal{Q}}_{\mathrm{V}}}$ of the proposed vision based method to predict the beam coherent time (VPBCT) under different training epochs. Thus, Fig.~15 of \cite{WXu} needs to be corrected as Fig.~6. Since Fig.~15 of \cite{WXu} is only utilized to demonstrate the convergence of the DNN of VPBCT and is not related to the performance of the VBPCT, this correction does not affect any conclusions related to VPBCT.

\balance

\end{document}